\documentclass[%
 reprint,
 amsmath,amssymb,
pra,
]{revtex4-1}

\usepackage{graphicx}
\usepackage{dcolumn}
\usepackage{bm}

\usepackage{upgreek}
\usepackage[dvipsnames]{xcolor}
\usepackage{amsmath}
\usepackage[utf8x]{inputenc}
\usepackage{braket}
\usepackage[english]{babel}

\newcommand{\mfzero}{$m_F = 0$\space}

\newcommand{\clocktransitionsecond}{$F=3,m_F=0\rightarrow F=2,m_F=0$\space}
\newcommand{\pumptransitionfirst}{$F=4\rightarrow F=4$\space}
\newcommand{\pumptransitionsecond}{$F=3\rightarrow F=3$\space}

\begin{document}

\title{Simultaneous two initial clock states preparation for thulium optical clock}

\author{E.\,Fedorova}
\email{fedorovaes@lebedev.ru}
\affiliation{P.N.\,Lebedev Physical Institute, Leninsky prospekt 53, 119991 Moscow, Russia}

\author{A.\,Golovizin}       
\affiliation{P.N.\,Lebedev Physical Institute, Leninsky prospekt 53, 119991 Moscow, Russia}

\author{D.\,Tregubov}
\affiliation{P.N.\,Lebedev Physical Institute, Leninsky prospekt 53, 119991 Moscow, Russia}

\author{D.\,Mishin}
\affiliation{P.N.\,Lebedev Physical Institute, Leninsky prospekt 53, 119991 Moscow, Russia}

\author{D.\,Provorchenko}
\affiliation{P.N.\,Lebedev Physical Institute, Leninsky prospekt 53, 119991 Moscow, Russia}

\author{V.\,Sorokin}
\affiliation{P.N.\,Lebedev Physical Institute, Leninsky prospekt 53, 119991 Moscow, Russia}
\affiliation{Russian Quantum Center, Bolshoy Bulvar 30,\,bld.\,1, \\Skolkovo IC, 121205 Moscow, Russia}

\author{K.\,Khabarova}
\affiliation{P.N.\,Lebedev Physical Institute, Leninsky prospekt 53, 119991 Moscow, Russia}
\affiliation{Russian Quantum Center, Bolshoy Bulvar 30,\,bld.\,1, \\Skolkovo IC, 121205 Moscow, Russia}

\author{N.\,Kolachevsky}
\affiliation{P.N.\,Lebedev Physical Institute, Leninsky prospekt 53, 119991 Moscow, Russia}
\affiliation{Russian Quantum Center, Bolshoy Bulvar 30,\,bld.\,1, \\Skolkovo IC, 121205 Moscow, Russia}
\date{today}

\begin{abstract}
Due to the low sensitivity of the thulium optical clock to black-body radiation and good accuracy and stability estimations, it appears to be one of the most promising transportable optical clocks. One of the leading systematic effects for Tm clock transition, namely the second-order Zeeman effect, can be canceled by probing two clock transitions between different hyperfine levels of ground and metastable states during the clock operation. We prepare the atoms in $m_F=0$ state of both ground hyperfine levels simultaneously and excite alternately two clock transitions. Here we demonstrate efficient optical pumping into both target states via single-frequency radiation at 418.8\,nm. The resulting population of $m_F=0$ states of ground hyperfine levels is 36\% and 3.8\% of the initial number of atoms with less than 4\% and 0.4\% on non-zero magnetic sublevels, correspondingly. We performed numerical simulations of the optical pumping process which is able to explain experimental results reasonably well.
\end{abstract}

\keywords{}
\maketitle
\section{Introduction}
\label{Section:intro}
For the past decades, tremendous progress within the field of optical clocks has allowed the development of a broad range of applications that are taking shape with optical clocks. Today the state-of-the-art optical clocks have reached unrivaled inaccuracy and instability of the low $10^{-18}$ level \cite{McGrew2018a,Nicholson2015a, Brewer2019a} employed for relativistic geodesy and gravimetry \cite{Takano2016,Grotti2018,Takamoto2020}, observations of cosmic radio sources with very-long baseline interferometry (VLBI) \cite{Clivati2017}, search for dark matter \cite{Derevianko2014} etc. To ensure further progress in this fields it is necessary to commit the transition from complex and bulky laboratory experimental setups to more compact and power-efficient systems, maintaining at the same time high stability and accuracy. It explains the fact that the development of transportable optical clocks is currently pursued in many laboratories worldwide \cite{Cao2017, Koller2017, Kong2020, Poli2014} and a number of projects are planned \cite{Hannig2019,Delehaye2018}. 

Besides operation reliability, it is necessary to ensure the high metrological performance of the transportable clocks in critical environments, such as application in the field. Thus the low sensitivity to the environment, including black-body radiation (BBR), is essential. In this perspective ion clocks, like  Lu$^+$ \cite{Arnold2018}, $^{229}$Th with an isomeric nuclear transition \cite{Campbell2012}  are of a great interest. As for the optical lattice clocks,  although they benefit from the high number of atoms and, hence, reduced fractional instability, as a rule, their sensitivity to BBR is quite high.

Among different atomic species suitable for optical lattice clock thulium has significantly lower sensitivity to the external static electric fields and black-body radiation \cite{Golovizin2019}, combining benefits of both ion and lattice optical clocks. The leading systematics here appears to be the second-order Zeeman shift due to relatively small hyperfine splitting (HFS) of both ground and upper clock levels. 

\begin{figure*}
\center{
\resizebox{0.95\textwidth}{!}{
\includegraphics{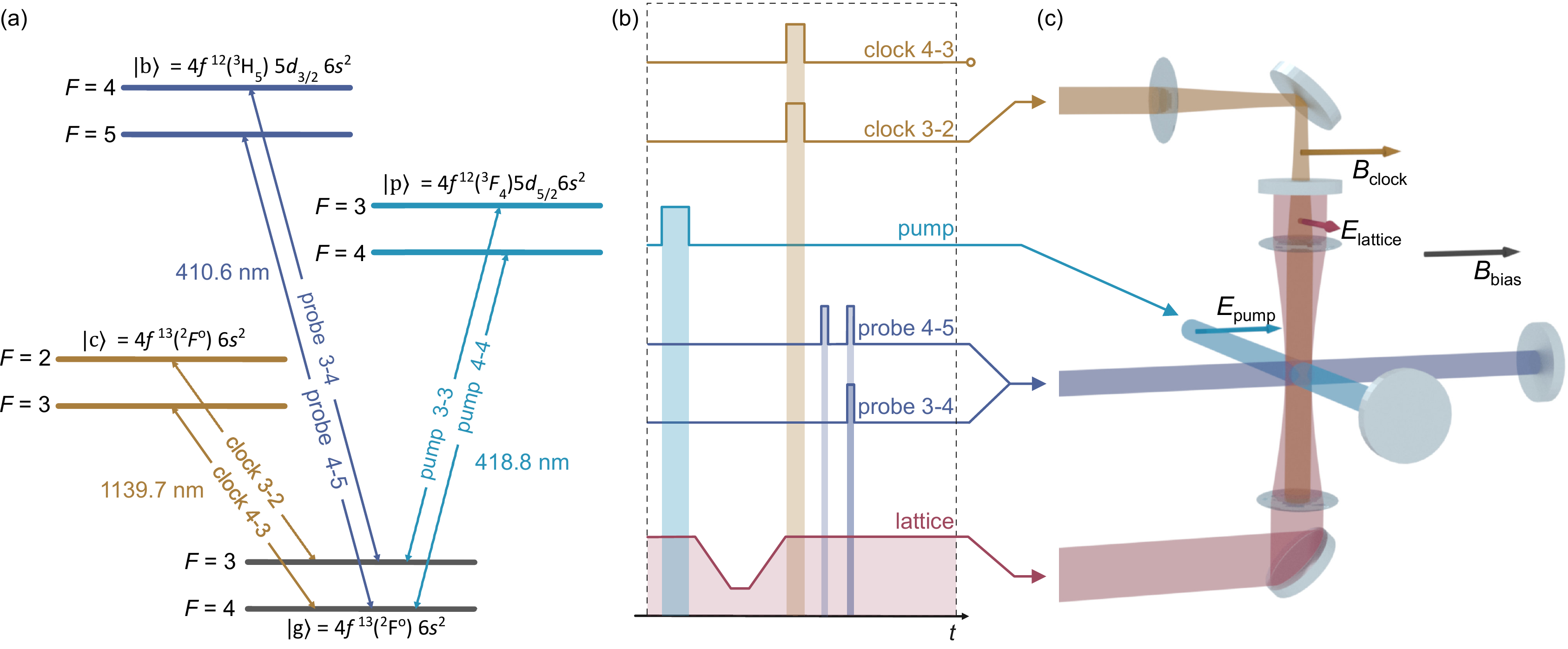}
}
\caption{(a) Energy levels and relevant transitions for neutral Tm. (b) Pulse sequence for clock transitions spectroscopy. After cooling and trapping into the optical lattice, atoms are pumped by a 5\,ms pulse of 418.8\,nm radiation into the $\ket{g, F=4, m_F=0}$ and the $\ket{g, F=3, m_F=0}$ states. 
The pumping is followed by a 55\,ms lattice depth ramp to remove atoms that are in higher-lying motional states. 
Then the $\ket{g, F=4} \rightarrow \ket{c, F=3}$ or $\ket{g, F=3} \rightarrow \ket{c, F=2}$ clock transition is excited by a corresponding 1\,ms clock pulse. 
After a 2\,ms delay a 0.2\,ms-long pulse $\ket{g, F=4} \rightarrow \ket{p, F=5}$ at $410.6$\,nm reads out the number of atoms remained in the $\ket{g, F=4}$ state and removes them from the trap. If the $\ket{g, F=3} \rightarrow \ket{c, F=2}$ clock transition was excited, 
7\,ms later the number of atoms in the $\ket{g, F=3}$ state is measured by two simultaneous $0.2$\,ms-long $\ket{g, F=3} \rightarrow \ket{p, F=4}$ and $\ket{g, F=4} \rightarrow \ket{p, F=5}$ probe pulses at $410.6$\,nm. 
(c) Sketch of the experimental setup. The bold arrows indicate the directions of the light beams linear polarization and the direction of the bias magnetic field.}
\label{fig:scheme}}
\end{figure*}

In general, there are two approaches to manage the clock transition frequency shift. The straightforward one consists in careful calculation of the shift and consequent frequency correction \cite{Lu2020,Huntemann2016a}. Alternatively, the frequency shift can be eliminated, for example by either interrogation of two clock transitions \cite{Burt2010, Yudin2011} 
or the frequency averaging over the clock transition sublevels \cite{Kaewuam2020, Dube2005, Oelker2019}.  In particular, for the second-order Zeeman shift in thulium the two hyperfine components of the clock transition $\ket{g,F=4, m_F=0} \rightarrow \ket{c,F=3, m_F=0}$ and $\ket{g,F=3, m_F=0} \rightarrow \ket{c,F=2, m_F=0}$ (hereinafter notation $\ket{g}$ refers to the ground state and $\ket{c}$ to clock state (Fig.\ref{fig:scheme}a); $F$ is the total momentum and $m_F$ -- the projection of total momentum) possess equal but opposite in sign quadratic Zeeman coefficients. 
Taking into account that polarizabilities of the $\ket{g,F=4, m_F=0}$ and $\ket{g,F=3, m_F=0}$ states, as well as $\ket{c,F=3, m_F=0}$ and $\ket{c,F=2, m_F=0}$ states, are pairwise equal, it is possible to construct a ``synthetic'' frequency \cite{Yudin2016} $\nu_s = (\nu_{43} + \nu_{32})/2 $, that is insensitive to both first- and second-order Zeeman shifts. Here $\nu_{43}$ ($\nu_{32}$) is the frequency of the $\ket{g,F=4, m_F=0} \rightarrow \ket{c,F=3, m_F=0}$ ($\ket{g,F=3, m_F=0} \rightarrow \ket{c,F=2, m_F=0}$) transition. By simultaneous interrogation of the two clock transitions, the contributions of the bias magnetic fields as well as the impact of the magnetic field fluctuations should cancel out, making the thulium clock entirely insensitive to the linear and quadratic Zeeman effect. However, to probe the two clock transitions during the clock operation cycle one needs to synchronously prepare atoms in both initial clock states.

In this work, we describe a simple approach to the simultaneous preparation of both initial clock states and demonstrate the excitation of the $\ket{g,F=3, m_F=0} \rightarrow \ket{c,F=2, m_F=0}$ component of the clock transition in thulium atoms.

The manuscript is organized as follows. In Section \ref{Section:418_HFS}, we discuss the transition at 418.8\,nm chosen for optical pumping. In Section \ref{Section:Experimental_procedure}, the experimental procedure of optical pumping and clock transitions interrogation is presented. Section \ref{Section:Pumping} describes an investigation of the optical pumping parameters and the simulation of the pumping process. Conclusions are summarized in Section \ref{Section:conclusion}.

\section{Transition for optical pumping}
\label{Section:418_HFS}

Optical pumping into a \mfzero state, which both initial clock states are, can be realized by driving a transition between levels with the equal total momentum $F=F'$ with $\pi$-polarized light since the transitions between $m_F =m_F'=0$ sublevels of those levels are forbidden \cite{Fedorova2019}. The nuclear spin of the only stable $^{169}$Tm isotope is $I=1/2$, so optical pumping into both initial clock states can be implemented using a $\ket{J=7/2} \rightarrow \ket{J=7/2}$ type transition, that comprises both $\ket{F=4} \rightarrow \ket{F=4}$ and $\ket{F=3} \rightarrow \ket{F=3}$ hyperfine components for two pumping procedures.
After laser cooling in the MOT, thulium atoms populate only the $\ket{g,F=4}$ state. Despite that, after excitation of the $\ket{F=4} \rightarrow \ket{F=4}$ transition, they decay with the probability of 1/35 to the $\ket{g,F=3}$ level, so both pumping procedures can be implemented simultaneously.

\begin{figure}
\center{
\resizebox{0.49\textwidth}{!}{
\includegraphics{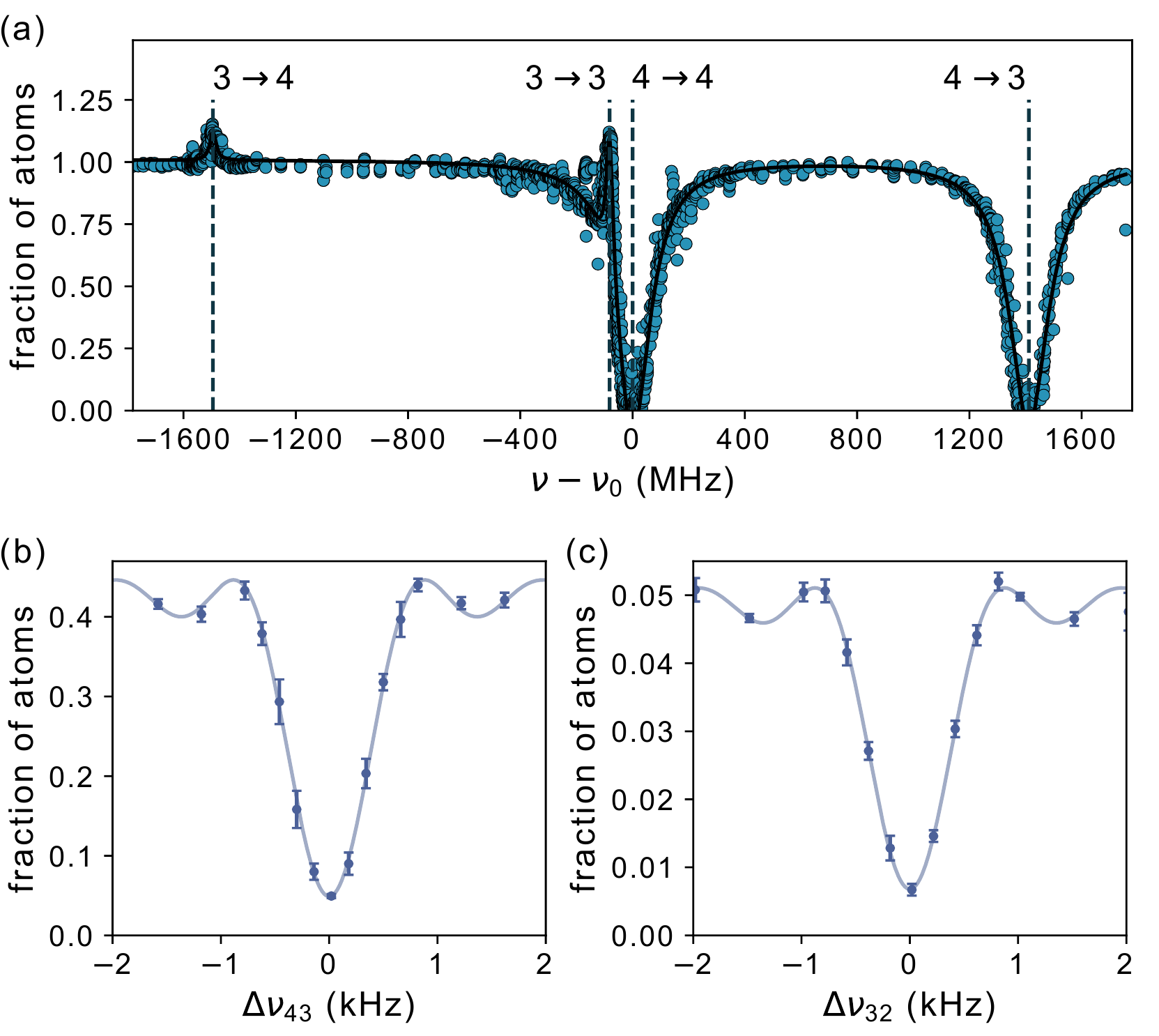}
}
\caption{a) Spectrum of the $\ket{g}\rightarrow \ket{p}$ transitions at 418.8\,nm in MOT. The frequency of the  $\ket{g,F=4}\rightarrow \ket{p,F=4}$ hyperfine component is $\nu_0=715\,701\,100$\,MHz. The dashed lines show the center frequencies of the other hyperfine components, the values of the hyperfine splittings are summarized in the Table \ref{tab:HFS}.
b) The absorption spectrum of the $\ket{g,F=4, m_F=0}\rightarrow \ket{c,F=3, m_F=0}$ clock transition.
c) The absorption spectrum of the $\ket{g,F=3, m_F=0}\rightarrow \ket{c,F=2, m_F=0}$ clock transition.}
\label{fig:418spectrum}}
\end{figure}

Here we use the $\ket{J=7/2} \rightarrow \ket{J=7/2}$ transition with the wavelength of  418.8\,nm  and the natural linewidth of $\Gamma_{p} = 2\pi \times 10$\,MHz (see Fig.\,\ref{fig:scheme}a) for optical pumping. To determine frequencies of the $\ket{F=4} \rightarrow \ket{F=4}$ and $\ket{F=3} \rightarrow \ket{F=3}$  hyperfine components, we measure the 418.8\,nm transition spectrum in the magneto-optical trap (MOT) (Fig.\,\ref{fig:418spectrum}a). We use the frequency-doubled cw Ti:Sa laser at 838\,nm as the light source, but it also can be replaced by an ECDL with or without frequency doubling to simplify the experimental setup. 
We detect the number of atoms in the $\ket{g, F=4}$ state
versus the detuning of the 418.8\,nm light. 
The detailed description of the thulium MOT and the detection procedure is given 
in \cite{Vishnyakova2014}.

To associate the spectrum features with the particular hyperfine components, one needs to consider the steady-state level occupation in the MOT.
In the thulium MOT only atoms in the $\ket{g, F=4}$ state are trapped. 
The small fraction of atoms is leaking to the $\ket{g,F=3}$ state due to non-resonant scattering, which results in trap losses and decreases the steady-state number of atoms in the MOT. 
Excitation of the $\ket{g, F=4} \rightarrow \ket{p, F=4}$ or the $\ket{g, F=4} \rightarrow \ket{p, F=3}$ transition leads to additional heating and losses, which result in two dips in the spectrum. 
In contrast, driving the $\ket{g, F=3} \rightarrow \ket{p, F=4}$ or the $\ket{g, F=3} \rightarrow \ket{p, F=3}$ transition acts as a repumper and increases the signal of luminescence. 

It is worth noting, that in the spectrum in Fig.\,\ref{fig:418spectrum}a
we cannot distinguish between the $\ket{g, F=4} \rightarrow \ket{p, F=4}$ and the $\ket{g, F=4} \rightarrow \ket{p, F=3}$ transitions, as well as between the $\ket{g, F=3} \rightarrow \ket{p, F=3}$ and the $\ket{g, F=3} \rightarrow \ket{p, F=3}$ ones.
However, observation of the efficient optical pumping into the $\ket{g,F=4,mF=0}$ sublevel (as discussed in the next section) when driving the transition associated with the third from the left feature in the Fig.\,\ref{fig:418spectrum}a allowed us to identify it as the $\ket{F=4} \rightarrow \ket{F=4}$ transition with certainty. After that, all other
hyperfine components can be unambiguously labeled, as shown in the figure.
From the measured spectra we calculate hyperfine splittings of both ground and upper levels of the 418.8\,nm pumping transitions, with results summarized in Table\,\ref{tab:HFS}. The absolute value of the hyperfine splitting of the upper pumping level $\ket{p}$, as well as the one of the ground state $\ket{g}$, is in agreement with previously reported values. However, the sign of the hyperfine splitting of the $\ket{p}$ level is found to be opposite to one previously reported in \cite{kolachevsky2007blue}.

One can see from the figure that the $\ket{F=4} \rightarrow \ket{F=4}$ and the $\ket{F=3} \rightarrow \ket{F=3}$ resonances overlap.
Indeed, the hyperfine splitting of the $\ket{p}$ level is close to the hyperfine splitting of the ground state $\ket{g}$: they differ only by $85$\,MHz\, which corresponds to $8.5$ $\Gamma_{p}$. As a result, two optical pumping procedures involving the $\ket{g,F=4} \rightarrow \ket{p,F=4}$ and the $\ket{g,F=3} \rightarrow \ket{p,F=3}$ transitions can be implemented by a single laser beam driving both transitions simultaneously. 
This approach, implemented in the work, significantly simplifies the setup which is important for compaction of the transportable clock.
The other transition in thulium that can be used for the simultaneous optical pumping by a single beam is one at the wavelength of $374.4$\,nm with the natural linewidth of $\Gamma'_{p} = 2\pi \times 16$\,MHz. In this case, the difference of hyperfine splittings is $202$\,MHz\,$ (13$ $\Gamma'_{p}$), which is also relatively small \cite{Brandt1977}.

\begin{table}[]
    \centering
\begin{tabular}{c|c|c}
    Source & HFS$_{\ket{g}}$, MHz & HFS$_{\ket{p}}$, MHz\\
   \hline \hline
   in MOT & $-1493.6\,(0.8)^\textrm{stat}(5)^\textrm{syst}$ & $-1413.6\,(1.5)^\textrm{stat}(5)^\textrm{syst}$ \\
        & $-1495.4\,(1.6)^\textrm{stat}(5)^\textrm{syst}$ & $-1411.8\,(1.0)^\textrm{stat}(5)^\textrm{syst}$ \\
    \hline
    mean & $-1494.5\,(1.5)^\textrm{stat}(5)^\textrm{syst}$ & $-1412\,(1.5)^\textrm{stat}(5)^\textrm{syst}$ \\
    \hline
    Literature & $-1496.550(1)$ \cite{van1980high}  & $+1411.0(7)$  \cite{kolachevsky2007blue}
\end{tabular}
\caption{Hyperfine splitting of the ground $\ket{g}$ and the upper pumping $\ket{p}$ levels.}
    \label{tab:HFS}
\end{table}

\section{Preparation of initial states  and interrogation of clock transitions}
\label{Section:Experimental_procedure}

\begin{figure*}
\center{
\resizebox{0.95\textwidth}{!}{
\includegraphics{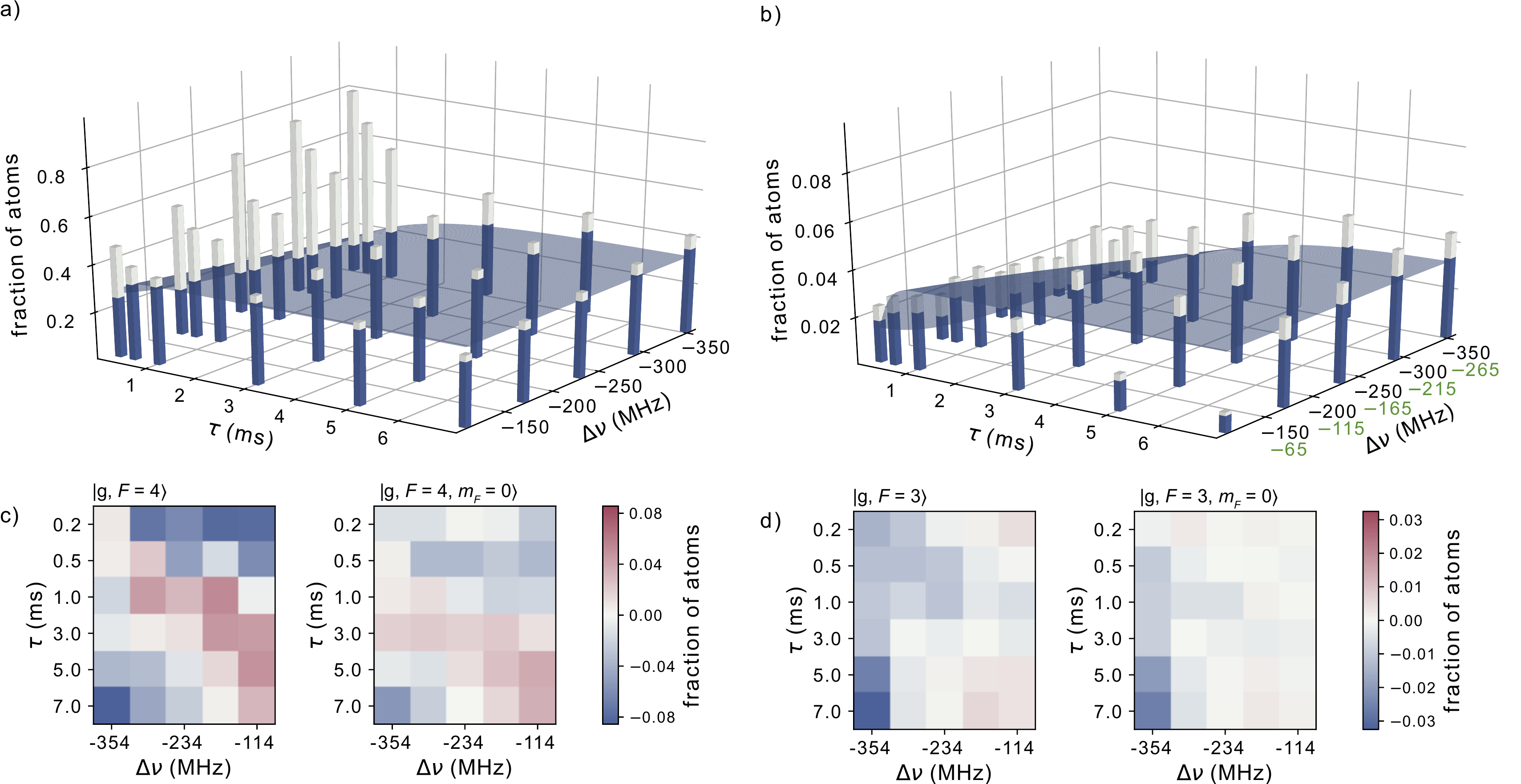}
}
\caption{
The efficiency of optical pumping to the $\ket{g, F=4, m_F=0}$ (a) and the ($\ket{g, F=3, m_F=0}$) (b) sublevels versus duration and detuning of the pumping-radiation pulse. The bars show the experimental data; the surface displays the calculation result based on the model described in Appendix \ref{Section:simulations}.
The total height of the bars corresponds to the fraction of the atoms in the $\ket{g, F=4}$ ($\ket{g, F=3}$) state. The height of the blue sub-bars illustrates the fraction of the atoms at the $\ket{g, F=4, m_F=0}$ ($\ket{g, F=3, m_F=0}$) sublevel. The green scale in (b) indicates the detunings in respect to the frequency of the $\ket{g,F=3} \rightarrow \ket{p,F=3}$ transition resonance and is shifted by 85 MHz from the detunings given in respect to the $\ket{g,F=4} \rightarrow \ket{p,F=4}$ transition frequency.
c) The discrepancy between the experimental points and the fit result for the fraction of the atoms in the $\ket{g, F=4}$ (left panel) and in the $\ket{g, F=4, m_F=0}$ sublevel (right panel). d)  The discrepancy between the experimental points and the fit result for the fraction of the atoms in the $\ket{g, F=3}$ (left panel) and in the $\ket{g, F=3, m_F=0}$ sublevel (right panel).}

\label{fig:pumping}}
\end{figure*}

Optical pumping and clock transitions excitation are performed for atoms trapped into the optical lattice formed by the  $1064$\,nm radiation. Although it is not exactly a magic wavelength, the differential polarizability of the clock transitions is sufficiently small for the chosen configuration (lattice beam with linear polarization orthogonal to the bias magnetic field) \cite{Golovizin2019} which enables observation of narrow, Fourier-limited, clock transition spectra.

The experimental procedure of pumping and clock transitions interrogation is similar to one described in  \cite{Golovizin2019}. The pulses sequence is shown in Fig. \ref{fig:scheme}b and the relative arrangement of the beams, the orientation of their polarization and the magnetic field is illustrated in Fig. \ref{fig:scheme}c. The optical lattice operates continuously. The cooling cycle lasts for 700 ms, producing the cloud of $10^6$ atoms with the temperature of $14$\,$\mu$K. After that, the laser beams and the magnetic field of the MOT are switched off, and about 40\% of atoms stay trapped in the optical lattice. Then we turn on the bias magnetic field  $B_0=0.5$\,G determining the direction of the quantization axis in the system.
After the magnetic field is on, we apply a pumping pulse at $418.8$\,nm (power $0.45$\,mW, detuning $\Delta \nu =174$\,MHz, duration $\tau =5$\,ms).
Here and after, the detuning of the pumping light is given in respect to the frequency of the $\ket{g,F=4} \rightarrow \ket{p,F=4}$ transition resonance.
The pumping beam is combined with its back reflection. The polarization of the pumping beam is linear and aligned with the direction of the magnetic field in order to excite $\pi$-transitions in atoms.

The pumping procedure causes heating of the atoms in the lattice, which may decrease the clock transition resonance contrast \cite{Blatt2009}. To remove atoms from higher-lying motional states, we linearly ramp the lattice depth down to 40\% of its initial level and back for 55\,ms (see Fig.\,\ref{fig:scheme}b). 
Then the $\ket{g, F=4} \rightarrow \ket{c, F=3}$ or the $\ket{g, F=3} \rightarrow \ket{c, F=2}$ clock transition is excited by a $1$\,ms \ $\pi-$ pulse and the corresponding detection procedure described below is performed.

The states populations after excitation of the clock transitions are measured by the fluorescence detection using the $\ket{g} \rightarrow \ket{b}$ transition at the wavelength of $410.6$\,nm (Fig. \ref{fig:scheme}a). 
The excitation of the $\ket{g, F=4} \rightarrow \ket{c, F=3}$ clock transition is followed by a $0.2$\,ms long $\ket{g, F=4} \rightarrow \ket{b, F=5}$ probe pulse aimed at measuring of the non-excited atoms at the $\ket{g, F=4}$ state.
The excitation of the $\ket{g, F=3} \rightarrow \ket{c, F=2}$ clock transition is followed by a $0.2$\,ms long $\ket{g, F=4} \rightarrow \ket{b, F=5}$ probe pulse, which removes atoms from the $\ket{g, F=4}$ level. Subsequently, we apply a $0.2$\,ms long $\ket{g, F=3} \rightarrow \ket{b, F=4}$ pulse, which pumps atoms from $\ket{g, F=3}$ to $\ket{g, F=4}$. A simultaneous $0.2$\,ms long probe pulse excites the $\ket{g, F=4} \rightarrow \ket{b, F=5}$ transition, and thus the number of non-excited atoms in the $\ket{g, F=3}$ state is measured.  
The spectra of the two clock transitions are shown in Fig. \ref{fig:418spectrum}(b,c). 
Note that the direct laser excitation of the $\ket{g, F=3} \rightarrow \ket{c, F=2}$ 1.14\,$\mu$m transition in thulium is demonstrated for the first time in this work.

\section{Pumping efficiency}
\label{Section:Pumping}
To determine the optimal parameters of the optical pumping we investigate its efficiency in dependence on the duration and the detuning of the 418.8\,nm pulse. The results are shown in Fig. \ref{fig:pumping}(a,b). To characterize the pumping efficiency, we measure the following values after the optical pumping procedure: 1) $\eta_4$ -- the fraction of atoms in the $\ket{g, F=4}$ state (the total heights of the bars in Fig.\ref{fig:pumping}a),
2) $\eta_{4,0}$ -- the fraction of atoms in the $\ket{g, F=4, m_F=0}$ state (blue sub-bars in Fig.\ref{fig:pumping}a), 3) $\eta_{3}$ -- the fraction of atoms in the $\ket{g, F=3}$ state (the total heights of the bars in Fig.\ref{fig:pumping}b), and 4) $\eta_{3,0}$ -- the fraction of atoms in the $\ket{g, F=3, m_F=0}$ state (blue sub-bars in Fig.\ref{fig:pumping}b). The fractions of atoms are considered in relation to the total number of atoms in the lattice without the optical pumping procedure. The number of atoms in the $\ket{g, F=4, m_F=0}$ ($\ket{g, F=3, m_F=0}$) sublevels is inferred as the number of atoms excited by the resonant $\ket{g, F=4} \rightarrow \ket{c, F=3}$ ($\ket{g, F=3} \rightarrow \ket{c, F=2}$) $\pi$-pulse, that in turn is evaluated from the number of non-excited atoms (the light-grey sub-bars in Fig. \ref{fig:pumping}a(b)). 

The overall trend consists in the increase of the optical pumping efficiency with the rise of the pumping pulse duration and then leveling off at values of $\eta_{4,0} = 0.36$ ($\eta_{3,0} = 0.038$), which corresponds to $140 \times 10^3$ ($15 \times 10^3$) atoms in the $\ket{g, F=4, m_F=0}$ ($\ket{g, F=3, m_F=0}$) state. Similar behavior are observed with increasing intensity at the fixed pulse duration. Qualitatively such dependence is consistent with the solution of the Bloch equations (see description of the Simple model in Appendix \ref{Section:simulations}) and confirms the transfer of the atoms to the dark state.
However, the stationary populations, observed in the experiment, differ quantitatively from the solutions of the Bloch equations. The discrepancy is associated with the heating and the losses that occur because the atoms undergo many scattering cycles during optical pumping.

The other trait is that the stationary populations do not depend on the frequency of the pumping beam in the broad range of detuning except the region near the resonance value.
The usage of the red-detuned light in combination with its back reflection makes it more probable that photons will be absorbed from the beam directed opposite to the motion of atoms. 
This selective absorption ensures relatively low heating of atoms. As the resonance approaches, the scattering rate for photons from both beams increases. Although the optical molasses mechanism becomes more effective, however, since the Doppler temperature for the pumping transition (240 $\mu$K) is higher than the depth of the optical lattice (20\,$\mu$K), the resulting losses increase. 

For the quantitative description of the optical pumping, we carry out the Monte-Carlo simulations of the population dynamics.  We consider all the magnetic sublevels of the $\ket{g, F=4}$, $\ket{g, F=3}$, $\ket{p, F=4}$ and the $\ket{p, F=3}$ states interacted with the $\pi$-polarized pumping radiation. To take into account the losses due to photon-scattering-induced heating, we introduce an empirical parameter $n_{thr}$, describing the average number of the scattering events before the atom escape the lattice. The value of $n_{thr}$ is chosen so that the stationary populations of the $\ket{g, F = 4}$ level coincided with the experimental values. 
The model based on conjoined fitting the experimental data of population dynamics for the $\ket{g,F=4}$ level in total and the $\ket{g, F = 4, m_F = 0}$ sublevel solely (the surface in Fig. \ref{fig:pumping}a shows fit for the number of atoms in the $\ket{g, F = 4, m_F = 0}$ state) allows describing the optical pumping into the $\ket{g, F = 3}$ level and the $\ket{g, F = 3, m_F = 0}$ sublevel without free parameters (the surface in Fig. \ref{fig:pumping}b illustrates prediction for the $\ket{g, F = 3, m_F = 0}$ state).
The panels in Fig. \ref{fig:pumping} c (d) demonstrate the difference between the experimental data and the theoretical prediction for the $\ket{g, F=4}$ level and the $\ket{g, F=4, m_F=0}$ sublevel (the $\ket{g, F=3}$ level and the $\ket{g, F=3, m_F=0}$ sublevel). The effect of the losses increase near the resonance is not considered in the model, so the discrepancy is higher for the small detunings.  The detailed description of the simulations is given in Appendix \ref{Section:simulations}. 

The small discrepancy between theory and experiment confirms that our understanding of the pumping process is correct and allows concluding that the parameters found are the optimal conditions for the optical pumping. 
Note that the efficiency of the optical pumping is almost insensitive to the intensity and the detuning in the broad range of parameters. As a result, there is no need for intensity stabilization. It is also possible to stabilize the pumping light frequency using, for example, a wavelength meter Angstrom WSU, which provides the stability of better than 10\,MHz. 

\section{Conclusion}
\label{Section:conclusion}
We demonstrate the simultaneous preparation of the two initial clock states for thulium atoms using optical pumping via 418.8\,nm  transition.
The close hyperfine splitting values of the ground and the upper state of the pumping transition make it possible to effectively excite both hyperfine components simultaneously using a single laser beam.
The obtained populations of the $\ket{g, F=4, m_F=0}$ and the $\ket{g, F=4, m_F=0}$ initial clock states are 36\% and 3.8\% of the initial number of atoms, which corresponds to $140 \times 10^3$ and $15 \times 10^3$ atoms, respectively.  
The pumping procedure efficiency is investigated and showed to work well in a wide range of parameters. We also perform the numerical simulation that takes the atomic losses into account and describes the experimental results well.

The population transfer to the $\ket{g, F=3, m_F=0}$ sublevel allows the spectroscopy of the \clocktransitionsecond clock transition. The parallel optical pumping enables the simultaneous interrogation of the two clock transitions, which would allow the formation of the synthetic frequency that is expected to be completely insensitive to the first- and second-order Zeeman effect.

\section{Acknowledgments}

Authors acknowledge the support of RSCF grants \#19-12-00137.

\appendix*
\section{Simulations}
\label{Section:simulations}
Numerical simulation is performed for the full magnetic-level structure of the ground $\ket{g}$ and the upper pumping $\ket{p}$ states, 32 sublevels in total. 
At the starting point, all the atoms are considered to be in the $\ket{g, F=4}$ ground state equally distributed over all magnetic sublevels.
In the following, we consider only the $\ket{g, F=4} \rightarrow \ket{p,F=4}$ and the $\ket{g, F=3} \rightarrow \ket{p,F=3}$ $\pi-$polarized laser excitation with zero frequency detunings. 
The spontaneous decay rate of the upper $\ket{p}$ level $\Gamma = 2\pi \times 10$\,MHz is set to 1 in the models, thus the timescale is normalized to the upper-level lifetime $\tau_0 = 1/\Gamma = 16$\,ns.

The Rabi frequency for the individual transition $\ket{g, F, m_F} \rightarrow \ket{p,F, m_F}$ can be calculated as

\begin{multline}
    \Omega_{F,m_F} = \Omega_0 \sqrt{(2 F +1)(2 J + 1)}  \\ \times \braket{F,m_F;1,0|F,m_F} \begin{Bmatrix} F & 1 & F \\ m_F & 0 & m_F \end{Bmatrix},
\end{multline}

where $J=7/2$ is the orbital momentum of the $\ket{g}$ (and the $\ket{p}$) state, $\braket{\dots|\dots}$ is a Clebsch-Gordan coefficient and $\{\dots\}$ is a Wigner's 6-j symbol.
$\Omega_0=\Gamma \sqrt{S/2}$ is a two-level atom Rabi frequency with saturation parameter $S$.

\begin{figure}[t!]
{
\includegraphics[width=0.49\textwidth]{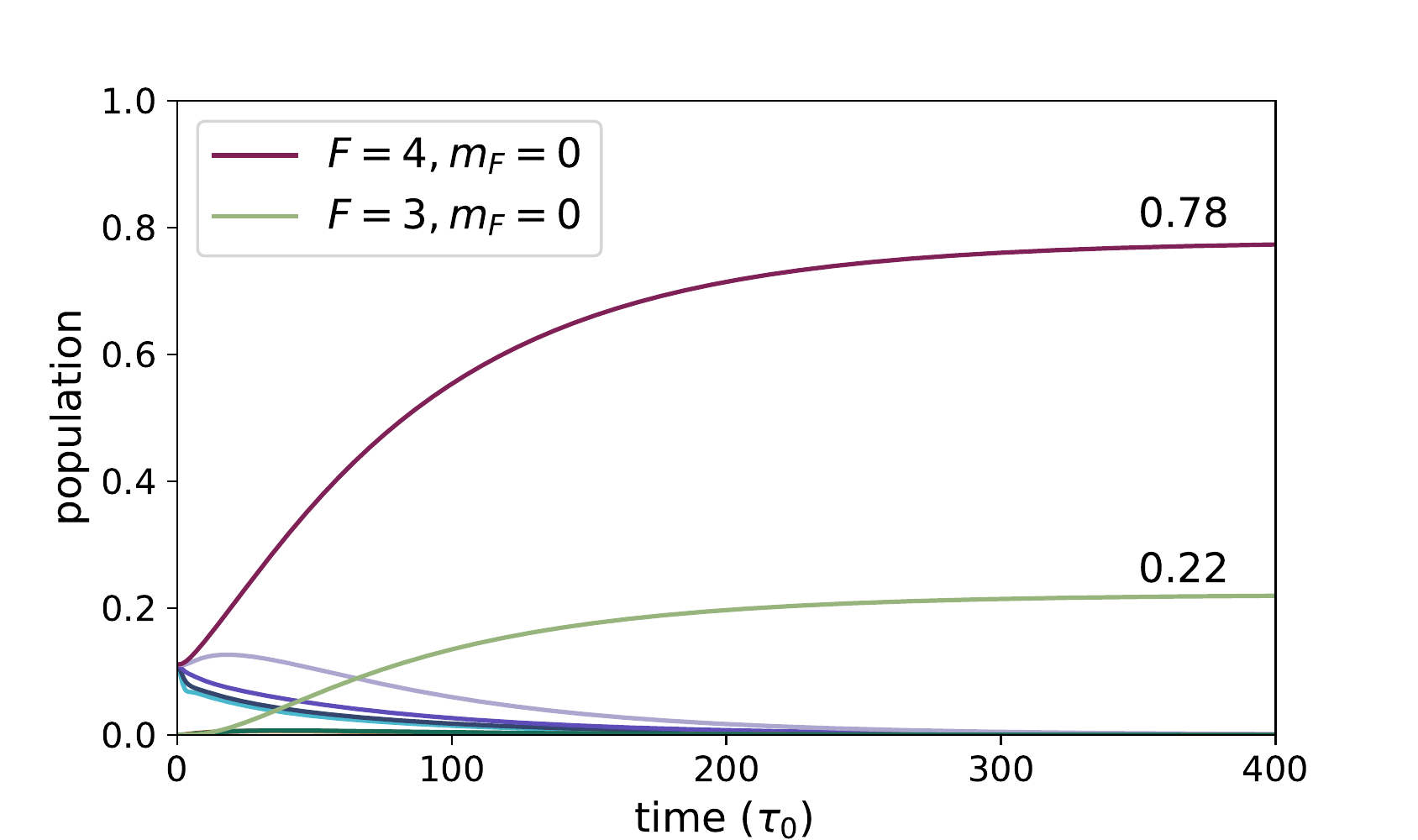}}
\caption{Evolution of the population of magnetic sublevels in the Simple model of optical pumping.}
\label{fig:simple_pumping}
\end{figure}

A Simple model implies solving the Bloch equation of population dynamics of a system density matrix, using QuTiP package \cite{qutip} without accounting for external motional degrees of freedom (for a detailed method's description see for example \cite{Fedorova2019}).
The simulated magnetic sublevels population dynamics of the $F=4$ and $F=3$ ground states is shown in Fig.\,\ref{fig:simple_pumping} with $m_F=0$ sublevels been labeled.
It is seen, that after $\tau = 300\tau_0 \approx 5\,\mu$s the steady state is reached for all atoms that have been pumped to the $\ket{F=4,m_F=0} (78\%)$ and $\ket{F=3,m_F=0} (22\%)$ states.

However, in the experiments described in Sec.\ref{Section:Pumping}, only $\eta_4\approx0.4$ of the initial number of atoms remain in the trap after the optical pumping in the $\ket{F=4}$ level and $\eta_3\approx0.04$ in the $\ket{F=3}$ level.
We associate the atoms losses with heating during optical pumping, particularly due to big recoil energy $E_r^{418} = 6.72$\,kHz from pumping photons at $418.8$\,nm  ($6.5$ times larger than the recoil energy $E_r^{1064}=1.04$\,kHz from the lattice photons) and typical experimental lattice depth of $100\,E_r^{1064}$.

In order to account for atoms losses during the pumping stage, we perform Monte-Carlo simulations of $10^5$ atoms' state evolution.
For every $j$-th evolution, an initial state $\psi^j[0] = \ket{F=4, m_F}$ of the atom is chosen randomly from uniform distribution over the $\ket{g, F=4}$ magnetic sublevels. 
The time evolution is divided into steps of $\tau_0$ length.
For each $i$-th step probability $\zeta^j[i]$ of a photon scattering is calculated as $\zeta^j[i] = \Omega_{F,m_F}^2 / (1 +2 \Omega_{F,m_F}^2)$ based on the instantaneous $F$ and $m_F$ values of the atom.
If the randomly generated number $\chi \in [0,1)$ is less then $\zeta^j[i]$, than photon scattering is assigned ($s^j[i]=1$), and the state of the atom $\psi^j[i+1]$ for the next iteration is chosen randomly between the ground state sublevels according to decay probabilities of the $\ket{p,F,m_F}$ level; otherwise $s^j[i]=0$ and $\psi^j[i+1] = \psi^j[i]$.
This procedure is repeated until an atom decayed to either the $\ket{g,F=4,m_F=0}$ or the $\ket{g,F=3,m_F=0}$ level.
Hereby, we trace the moments of spontaneous decay events as well as atom's state evolution.

After that, we perform the atoms' trajectory analysis to determine whether each atom was lost from the trap or not. 
To do this for each $j$-th atom a maximum number of scattered photon $n^j_{sc} = - n_{thr}\times ln(\chi^j)$ is generated, where $\chi^j \in [0,1)$ is a random number and $n_{thr}$ is some threshold number of scattered photons, that is varied to fit the experimental data.
If the number of scattered photons $\sum_i s^j[i]$ is less than $n^j_{sc}$, then we assign that j-th atom has reached the final state $\ket{g,F=4,m_F=0}$ or $\ket{g,F=3,m_F=0}$ at time $t_j$, otherwise, it is assigned to be lost from the level $\ket{g,F=4}$ or $\ket{g,F=3}$ at a time when $n^j_{sc}$-th scattering occurred.
If the state of the atom changes from the $\ket{g,F=4}$ to the $\ket{g,F=3}$ before the atom is lost from the trap or reaches the $\ket{g,F=4,m_F=0}$ state, we also assign this as the loss of the atom from the $\ket{g,F=4}$ level.

\begin{figure}
\resizebox{0.49\textwidth}{!}{
\includegraphics{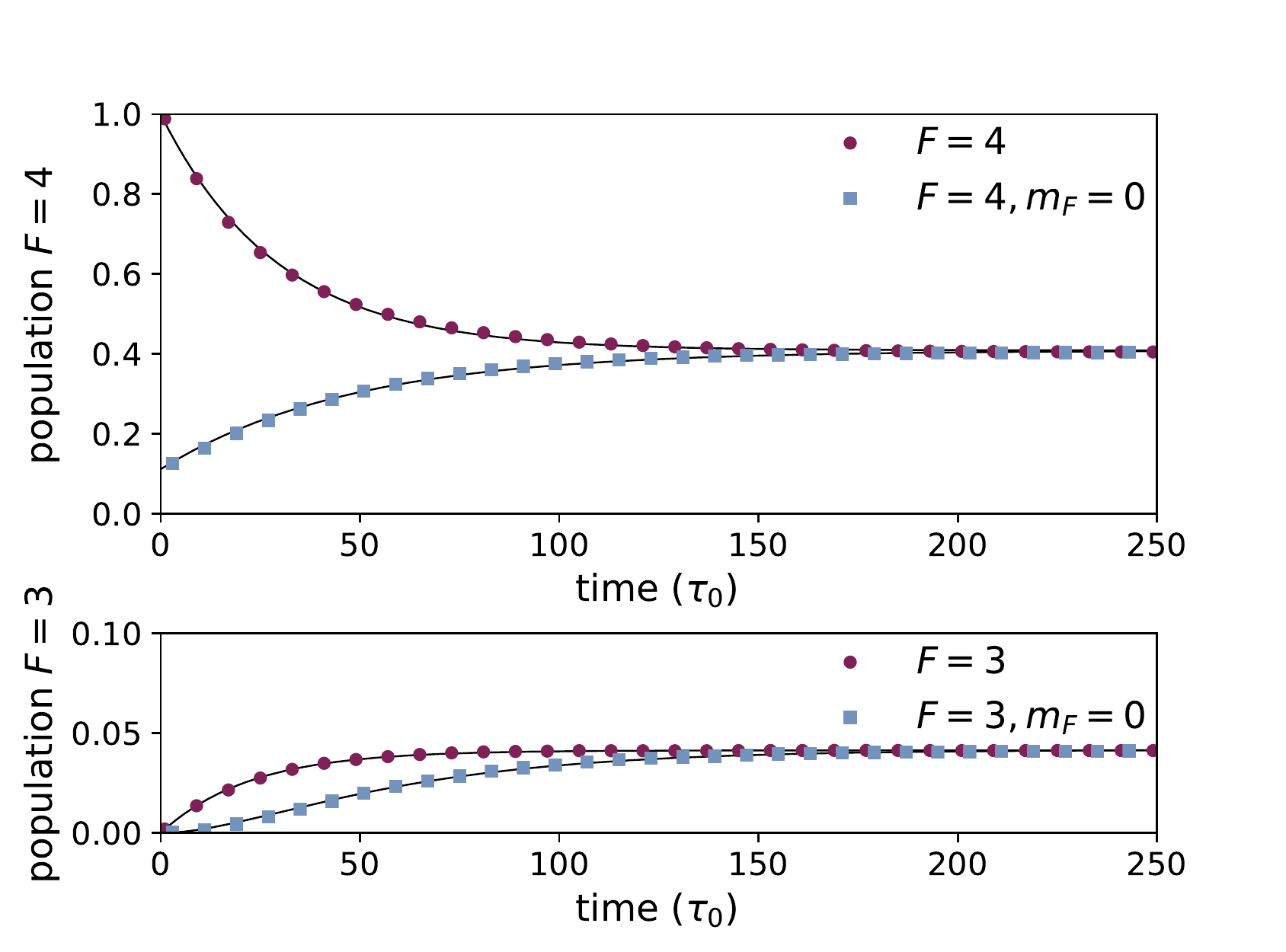}}
\caption{Evolution of the $\ket{g,F=4}$ (top panel) and the $\ket{g,F=3}$ (bottom panel) levels population.
Red circles indicate the fraction of atoms in the corresponding $F-$state, blue squares represent the fraction of atoms in the $\ket{m_F=0}$ sublevels. 
Solid lines are exponential fits \ref{eq:fit_functions1}-\ref{eq:fit_functions4}.
}
\label{fig:pumping_MC}
\end{figure}

Figure \ref{fig:pumping_MC} shows the results of calculation. 
Red circles represent the evolution of the number of atoms in the $\ket{g,F=4}$ (the top panel) and the $\ket{g,F=3}$ (the bottom panel) states.
The threshold number of scattered photons $n_{thr}$ is set to 8 to achieve the final fraction of atoms at the $\ket{g,F=4}$ level $\eta_4=0.41$ equal to that in the experiment.
Blue squares show the number of atoms in the $\ket{m_F=0}$ sublevel of the corresponding hyperfine levels.
In the experiment we do not observe full polarization of atoms in the $\ket{g,F=4}$ state achieving $\eta_{4,0}/\eta_4=0.9$. 
The possible reasons could be a non-perfect clock $\pi-$pulse, a difference of a real dark state from the $\ket{g,F=4,m_F=0}$ due to pump beam polarization impurity, and contribution from atoms in the $\ket{g,F=3}$ state when measuring the number of atoms in the $\ket{g,F=4}$ state.
From this simulation we infer the steady-state value of atoms' fraction in $\ket{g,F=3}$ level $\eta_3=0.042$.
This is in a very good agreement with $\eta_3^{exp}=0.045$ of the maximum fraction of atoms in the $\ket{g,F=3}$ state measured in the experiment.
The difference can be associated with losses during the $\ket{g,F=4}$ population readout and, of course, due to model simplifications.

\begin{figure}
\resizebox{0.49\textwidth}{!}{
\includegraphics{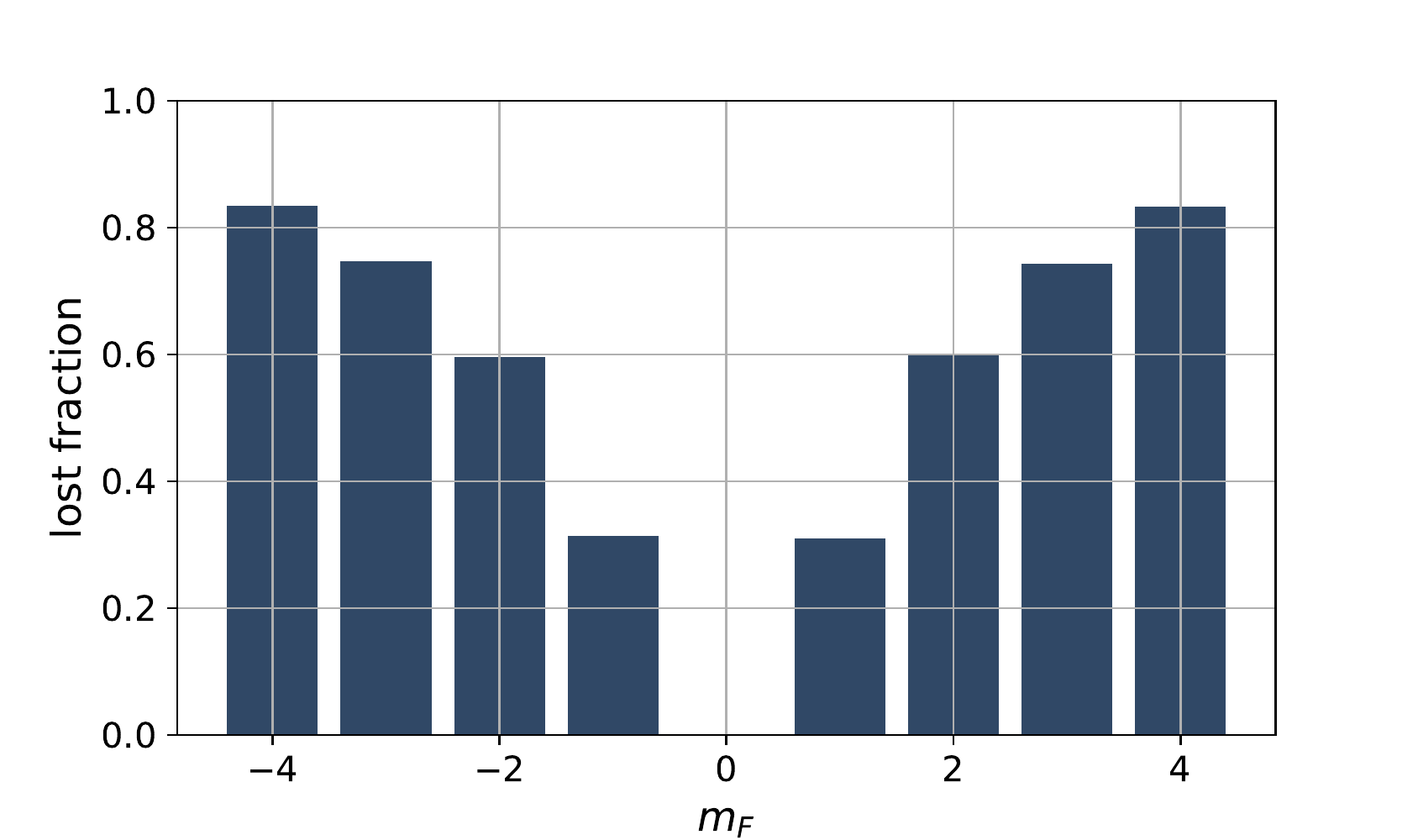}}
\caption{The fraction of atoms that are lost from the trap during optical pumping vs. the initial magnetic quantum number.}
\label{fig:losts}
\end{figure}

It is interesting to look at the distribution of atoms losses depending on the initial state $m_F$ as shown in Fig.\,\ref{fig:losts}.
As expected, the loss probability increases for the outer magnetic sublevels since the number of scattering events needed to reach the dark $m_F=0$ state also increases.
We note, that using a pre-pumping pulse of $\pi-$polarized radiation resonant with the $F=4\rightarrow F=5$ transition (i.e. the second-stage cooling transition) one can start with a more centered distribution of the atoms among the magnetic sublevels.
This can increase $\eta_4$ up to $0.72$ in our experiment, however at the same time reducing the number of atoms in the $\ket{g,F=3}$ state to $\eta_3\approx0.032$. 
This scheme may be used when the $\ket{g,F=3}$ state is not involved in the experiment, or with an additional microwave pulse to transfer fraction of atoms from the $\ket{g,F=4,m_F=0}$ to the $\ket{g,F=3,m_F=0}$ state~\cite{pershin2020microwave}.

To compare pumping performance obtained in theoretical calculations described above with the experimental data, we first fit levels' population dynamics (see Fig.\,\ref{fig:pumping_MC}, solid thin lines) using the following functions:

\begin{align}
\label{eq:fit_functions1}
    \eta_{4}(t) &= \eta^{sat}_4 + (1 - \eta^{sat}_4) e^{-t / \tau_4} \\
    \label{eq:fit_functions2}
    \eta_{4,0}(t) &= 1/9 + (\eta^{sat}_{4,0} -1/9)e^{-t / \tau_{4,0}} \\
    \label{eq:fit_functions3}
    \eta_{3}(t) &= \eta^{sat}_3 (1 - e^{-t / \tau_3}) \\
    \eta_{3,0}(t) &= \eta^{sat}_{3,0} (1 - e^{-t / \tau_{3,0}})^2
\label{eq:fit_functions4}
\end{align}

Here time of the evolution $t$ is in units of $\tau_0$. We obtained $\tau_4 = 29.3$, $\tau_{4,0} = 47.7$, $\tau_3 = 21.9$ and $\tau_{4,0} = 43.2$ and $\eta^{sat}_3=0.042$ (as is discussed earlier, we adjust the threshold of scattered photons $n_{thr}$ in a way to $\eta^{sat}_4$ match the experimental value of $0.41$).
The coefficient $1/9$ in Eq. (\ref{eq:fit_functions2}) comes from the assumption that initially atoms equally populate all 9 magnetic sublevels of $\ket{g,F=4}$ state.
The second power of expression in parentheses in Eq. (\ref{eq:fit_functions4}) reflects that the number of atoms in the $\ket{g,F=3}$ state simultaneously increases during optical pumping.

In the described model, we consider pumping radiation to be in the exact resonance with both transitions.
For each transition, the variation of the frequency detuning should only change the time scale of the pumping process inversely to photon scattering rate:
\begin{equation}
    \Gamma_{sc} = \frac{\Gamma}{2} \frac{S}{1+S+4(2 \pi \Delta\nu/\Gamma)^2}
\end{equation}
where $S$ is the saturation parameter and $\Delta\nu$ is the frequency detuning. 
In the following, we consider $\Delta\nu$ to be the frequency detuning of pump beam radiation from the \pumptransitionfirst transition since most of the photons are scattered on it. 

To fit the experimental data presented in Sec.\,\ref{Section:Pumping}, we use functions \ref{eq:fit_functions1}-\ref{eq:fit_functions4} while replacing $t$ with 
\begin{equation}
    n_{sc}(\tau,\Delta\nu) = a\times \tau \frac{\Gamma_{sc}}{\Gamma} = \frac{a\times \tau}{2} \frac{S}{1+S+4(2 \pi \Delta\nu/\Gamma)^2},
\end{equation}
where $a$ and $S$ are the free fit parameters, and $\tau$ is the pump pulse length.
Values of $a$, $S$, and $\eta^{sat}_{4,0}$ are determined from the list-square fit of the data presented in Fig.\,\ref{fig:pumping}a with functions \ref{eq:fit_functions1} and \ref{eq:fit_functions2}. 
Semi-transparent 2D-surface in Fig.\,\ref{fig:pumping}a shows obtained $\eta_{4,0}(\tau,\Delta\nu)$.
The semi-transparent 2D-surface in Fig.\,\ref{fig:pumping}b shows $\eta_{3,0}(\tau,\Delta\nu)$ with no fit parameters: values of $a$ and $S$ are used from above,  $\eta^{sat}_{3,0}=0.9\eta^{sat}_3=0.038$ is calculated from theoretical value of $\eta^{sat}_3$ and coefficient $0.9=\eta_{4,0}/\eta_4$.
One can see that theoretical curves describe the experimental data well. 
The major deviation occur for $\eta_{3,0}(\tau,\Delta\nu)$ for pump beam detuning of $-114$\,MHz. 
This can be explained due to the small detuning of pump radiation from \pumptransitionsecond transition of $-30$\,MHz, which increases heating and losses.

\bibliography{references}
\end{document}